\documentclass{mn2e}
\usepackage{psfig}

\def\ltsima{$\; \buildrel < \over \sim \;$}
\def\lsim{\lower.5ex\hbox{\ltsima}}
\def\gtsima{$\; \buildrel > \over \sim \;$}
\def\gsim{\lower.5ex\hbox{\gtsima}}
\newcommand{\f}{\frac}

\newcommand{\al}{\alpha}

\begin{document}
\title{Superbursts and long bursts as surface phenomenon of
compact objects.}

\author[ Sinha, Dey, Ray \& Dey]
{Monika Sinha $^{1,~ 2 ~\ddag}$, Mira Dey $^{1,~ 3 ~\dagger}$,
Subharthi Ray $^{4 \dagger\dagger}$ \& Jishnu Dey $^{2,~ 3~ \dagger}$\\
$^1$ Dept. of Physics, Presidency College, Calcutta 700
073, India\\ $^2$ Azad Physics Centre, Dept. of Physics, Maulana
Azad College, Calcutta 700 013, India\\ $^3$ Senior Associate,
IUCAA, Pune, India \\ $^4$ Theoretical Physics Division,
Physical Research Laboratory, Navrangpura, Ahmedabad 380 009,
India \\ $^{\dagger\dagger}$ e-mail:sray@prl.ernet.in \\
$\dagger$ permanent address; 1/10 Prince
Golam Md. Road, Calcutta 700 026, India;
e-mail:deyjm@giascl01.vsnl.net.in.\\
Work supported in part by DST grant no. SP/S2/K-03/2001, Govt. of
India.\\ $^\ddag$ CSIR Net Fellow. }

\maketitle

\begin{abstract}
  X-ray  bursts from  compact stars is believed to be due
to type I thermonuclear processes which  are short lived,
typically $\sim ~10 ~{\rm to} ~100~s$. There are  some low mass
X-ray binaries (LMXB) like 4U~1820$-$30, 4U~1636$-$53,
KS~1731$-$260 and Serpens X-1, known  as Super Bursters (SB)
which  emit X-rays close to the Eddington luminosity limit for
long periods of several hours. Recently  there are reports  of
some long bursters  (LB), which have bursts lasting 6-25 minutes
(Kuulkers et al. 2002b) whereas the 4U~1735$-$44 has a burst
period of 86 minutes (Cornelisse et al, 2000). The full
explanation of type I bursts in these stars is somewhat
problematic, in so far as bursts become less frequent and
energetic as the global accretion rates increase, as discussed by
Bildsten recently (2000).

    We suggest that these bursts from SB and LB may be due to
breaking and re-formation of diquark pairs, on the surface of
realistic strange  quark stars (ReSS). We use the beta
equilibrated u, d and s quark model of Dey et al. (1998) (D98)
and Li et al. (1999a  (Lia) and 1999b (Lib)) and allow for  spin
dependent hyperfine interaction between quarks. The interaction
produces pairing of specific colour-spin diquarks, leading  to
further lowering  of energy by several $MeV$-s for each pair, on
the average.

    Diquarks are expected to break up due to the explosion and shock
of the TN process. The subsequent production of copious diquark
pairing may produce sufficient energy to produce  the very long
bursts seen in SB or LB. We do not claim to be able to model the
complicated process in full. However the estimated total energy
liberated, $10^{42}$ ergs, can be explained in our model with the
calculated pair density $\sim ~0.275/fm^3$ and a surface
thickness of only half a micron, if the entire surface is
involved. The depth of the surface involved in the process may be
only few microns if the process is restricted to small part of
the surface near the equator as suggested by Bildsten (2000).

If SB and LB are surface phenomenon, then  recurrent superbursts
found by 4U~1636$-$53 by Wijnands (2001) at an interval of 4.7
year and the quick cooling of KS~1731$-$260(Kuulkers et al 2002b)
could be natural in this model (Wijnands et al. 2001 and 2002).
\end{abstract}


\section{Introduction}


It is intriguing to surmise that the elusive properties of some of
the most compressed objects in nature namely the compact stars,
showing superbursts, may be accounted for by the spin alignment of
pairs of the smallest components of matter, - namely the quarks.

Recently there has been lot of activity centered around the
possibility of lowering of the spin zero state of a diquark in
dense matter (see for example the review by Rajagopal \& Wilczek
(2000) and references therein). There has also been the suggestion
that diquarks may be present (Bhalerao \& Bhaduri 2000) like
droplets, i.e. with total negative energy rather than  just a
negative correlation energy as in a superconducting pair.

    The large $N_c$ expansion, for the number of colours $N_c$,
suggests a tree level mean field calculation for quark matter.
Using a realistic two quark potential within this scenario leads
to realistic strange stars (ReSS) which are self bound. The matter
has a minimum energy at a density which is high ($\sim 4$ to $5$
times the normal nuclear matter density, $\rho_0 ~=~ 0.17/fm^3$)
as shown in D98. We now estimate the spin correlations in this
matter, which is washed out in the mean field approximation of
D98, being a $1/N_c$ effect, by a simple perturbative calculation
using various sets of smeared spin-spin interaction which were
tested out for the isobar-nucleon mass difference in Dey and Dey
(1984).

    The importance of the exercise may be far-reaching, in so far
as there is a rich plethora of unexplained phenomena in the X-ray
emission pattern of compact stars. For example the  compact object
claimed to be ReSS (Lia), the SAX J1808.8$-$3658, show erratic
luminosity behaviour and a very long burst time (Wijnands et al.
2001b). The recent discovery of the compactness of
RXJ~1856.5$-$3754 also supports the possibility of strange stars
(Drake et al, 2002)

We suggest that the structure of the surface of the star may be as
important as the nature of the accretion disk variations in
explaining these phenomena.

    It is worth noting that according to Kapoor and Shukre
(2001), even radio pulsars are so compact that it is difficult to
explain their mass and radius from neutron star models. They
prefer ReSS.

\section{The astrophysical problem, Super Bursters.}

    Type~-~I X~-~ray bursts in LMXB systems are characterized by
fast rise times (of the order of seconds), long decay times
(seconds to minutes), spectral softening during the bursts, and
recurrence times of hours to days. In contrast, the physics
behind long lasting `super bursts' seen recently in several stars
is not yet well known, which is mostly a result of the very
recent discovery of such bursts and the limited information
available about them (Wijnands 2001). The first superburst was
reported by Cornelisse et al (2000) from  the LMXB 4U~1735$-$44
in 2000. Wijnands (2001) reported two superbursts for
4U~1636$-$53 and Heise et al. (2000) for KS 1731$-$4260 and
Serpens X-1. For 4U~1636$-$53 two clear superbursts have been
observed, although some of the smaller flares seen might also be
related to superburst phenomenon (Wijnands 2001).

    Spin alignment may be spoiled during
the prolonged strong accretion and the shock of the thermonuclear
bursts \footnote{or the conversion of the normal accreting matter
into strange matter if one prefers the other scenario for the
short initial burst, (Bombaci and Datta 2000)}. The realigning of
the spin zero diquarks could be a very natural scenario for the
superbursts, - which will be a slower process, since the u, d, s
quark and electron percentages are equilibrated with the beta
stability and charge neutrality conditions involving slower weak
and electromagnetic processes. The diquark energy lowering is a
strong process and the magnitude of energy release is of the same
order as that of a thermonuclear reaction (TR).

The mechanism for superburst that we suggest is outlined below :

compact stars with a high rate of accretion undergo thermonuclear
bursts lasting typically upto 20 seconds. During the high
accretion and the TR, the quark pairs (in particular the ud
pairs), - bound by the short range spin-spin interaction, - break.
After a sufficiently long time (expected to vary substantially
from star to star due to the statistical nature of the processes
and also the variation of the surface conditions {\footnote{This
time interval may be a few minutes [for example 6-25 minutes for
the 10 superbursts observed in GX 17+2, Kuulkers et al. (2002a)],
or several years [for example 4.7 years as in 4U 1636$-$53,
Wijnands 2001]}} - most of the pairs are broken and after a final
TR, the pairs start realigning.

The realigning of pairs will lead to a prolonged emission of
energy which may be transformed into X-rays leading to the
superbursts. This time may also vary for the same reasons as
above thus explaining the 86 min. superburst in 4U 1735$-$44
(Cornelisse et al. 2002), 4 hours in Serpens X-1 (Cornelisse et
al. 2000) and half a day in KS 1731$-$260(Kuulkers et al. 2002b).

According to this scenario there will be a link with the extreme
macro physics of compact stars of sizes of the order of kms and
masses of the order of solar masses with small diquarks paired by
a short range force of few $fm$ and bound by few $MeV$. There is
no time-scale limit in this model between two superbursts and one
may assume that the 4.7 years gap between the two superbursts
seen in 4U~1636$-$53, is the upper limit for the interval since
due to the erratic sampling of RXTE/ASM which detected these
bursts some intermediate bursts might have been missed or partly
recorded (Wijnands 2001).

    4U~1820$-$30 which was a candidate for ReSS in D98 also
shows superbursts lasting 3 hrs and a very interesting  model has
been proposed to explain this Cummings and Bildsten (2001),
Strohmayer and Brown (2001). These authors suggest that for this
particular star, which they {\it assume} to be a neutron star, the
superbursts are due to unstable carbon burning, the carbon being
possible remnants from the ashes of a Helium thermonuclear burst
buried deep down ($\sim~ 10$m) in an `ocean', mixed with iron.

    This is in sharp contrast to our scenario where we find enough
ud quark pairs, within depth of about $10^{-5}$ cm of the high
density star skin, to provide the energy of the burst (estimated
by Strohmayer and Brown (2001) to be $1.4\times 10^{42}$ ergs
equivalent to $10^{47}$ MeV). The strongest constraint according
to them on their scenario is that another such superburst should
not be detected within a time scale less than a decade. So, if
4U~1820$-$30 shows another superburst within the next few months
or years, the assignment of ReSS for this star D98 will find
additional support from present considerations.

    We thus find that our model admits of a rather attractive
alternate solution to the problem which is also applicable to the
other superbursters. It must be mentioned that Wijnands (2001) and
(TB) agrees that carbon burning is unlikely for 4U~1636$-$53 since
it seems to be a hydrogen-~ accreting source and carbon burning is
more likely for helium-~accreting sources.

    In the next following sections we present our model in some details.

\section{A brief introduction of the model}\label{sec:ref}

The quark (q) star model described in D98, which is also the same
model used here, is a realistic model of quark matter composed of
three flavours u,d and s as well as electrons. In hadron
spectroscopy,  using a potential model,  a realistic q-q
interaction contains asymptotic freedom (short range)  and
confinement (long range). However, in the case of quark matter,
confinement is softened by Debye screening which diminishes the
attractive long range part. The effect of this screening increases
with density so that deconfinement is further enhanced at high
densities.

Another very important consideration is the quark masses. The
general belief is that chiral symmetry tends to be restored at
high density which means quarks become lighter. The density
dependence of quark masses, therefore, is a reflection of the
chiral symmetry restoration (CSR in short) of QCD at high density
and can be alternatively represented  as a density dependence of
the strong coupling constant using simple Schwinger Dyson
techniques. We refer the interested reader to Ray et al, 2000. The
density dependence of quark masses , in this model, is taken care
of by the ansatz :
\begin{equation}
M_i = m_i + M_Q \,sech(\nu \f{\rho_B}{\rho _0}), \;\;~~~ i = u, d,
s.
\label{eq:qm}
\end{equation}
where $\rho_B = (\rho _u+\rho _d+\rho _s)/3$ is the baryon number
density, $\rho _0 = 0.17~fm^{-3}$ is the normal nuclear matter
density, and $\nu$ is a  parameter. At high $\rho_B$ the quark
mass $M_i$ falls from its constituent value $M_Q$ to its current
one $m_i$ which we take to be (D98): $m_u = 4 \;MeV,\; m_d = 7
\;MeV,\; m_s = 150 \;MeV$. $M_Q \sim 310 MeV$. Possible variations
of the CSR can be incorporated in the model through  $\nu $.

With these two ingredients ( along with the constraints of  $\beta
$ - equilibrium and charge neutrality) it is found  that energy
per baryon  is lower than that of $^{56}{\rm Fe}$ and has a
minimum at a density $\sim$ 4 to 5 times the normal nuclear
density $\rho_0$. This is a relativistic mean field calculation
with a central potential (screened Richardson potential)  where
only the Fock term contributes. Strange quark matter is thus self
bound by strong interaction itself. The energy density and
pressure of this matter lead to strange quark star through the TOV
equation with mass and radius depending on the central density of
the star.

   Equations of state obtained for two different values of $\nu $,
namely,  eos1 and eos2,   lead to different maximum masses of the
stars and their corresponding radii. (Table 1). Also given in
Table 1 is the energy/baryon of the strange quark matter to be
compared with that of $^{56}{\rm Fe}$.

The surface of the star starts at this high density of $\sim$ 4 to
5 times the normal nuclear density $\rho_0$.  The density inside
the star can be larger, the limit being $\sim$ 15 times at the
core when gravitational instability sets in. Thus at the surface
there are massive quarks (about 100 MeV for u,d and 250 MeV for
s-quarks) whereas at the centre of  a massive star with density
$\sim$ 10 to 15 times the normal density $\rho_0$ the masses
approach the current quark masses 4,7 and 150 MeV for u,d and s
respectively).

\begin{table}
\caption{Properties of  the maximum mass strange star
configuration obtained for different forms for CSR : $M_G$ is the
gravitational (maximum) mass, R is the corresponding radius,
$n_c$ the central number density, $\rho_c$ the central mass
density. Our EOS for different choices of the parameters are
denoted as follow: (eos1) $\nu = 0.333$, $\al_0 = 0.20$; (eos3)
$\nu = 0.333$, $\al_0 = 0.25$; (eos3) $\nu = 0.286$, $\al_0 =
0.20$. The reference for the binding per baryon B.E./A is 930.6
$MeV$ for $Fe^{56}.$}  \vskip 1cm
\begin{center}
\begin{tabular}{|c|c|c|c|c|c|}
\hline
EOS & $M_G$ & R & $n_c$ & $\rho_c$& B.E./A \\
&$(M_{\odot})$& (km)& $(fm^{-3})$ &$( 10^{14}g/cm^{3})$ &$MeV$\\
\hline
 eos1  & 1.437 & 7.06 & 2.324 &  46.90 & 888.8 \\
 eos3  & 1.410 & 6.95 & 2.337 &  48.19 & 844.6\\
 \hline
\end{tabular}
\end{center}
\end{table}
\vskip 0.5cm

\section{The spin-spin potential}

    The $\Delta $ isobar is an isospin 3/2  of the spin 3/2
excitation of the nucleon seen at about 1232 MeV.  To calculate
nucleon and isobar mass difference (of about 300 MeV) we need a
finite range spin spin interaction. Indeed,  the quark-quark
interaction has also a spin dependent component which can be
obtained either from one-gluon exchange between quarks or from the
instanton induced interaction. This part of the potential is of
delta function range which can be  transformed to a smeared
potential by introducing the idea of either a finite glue-ball
mass or a secondary charge cloud screening as in electron-physics
(Bhaduri et al, 1980).

The essential idea is to get a smeared Gaussian potential with a
renormalized strength. The smearing and the strength can be
obtained by fitting them  to observables like nucleon- $\Delta $
mass splitting  and the magnetic dipole transition from $\Delta $
to nucleon. We borrow the allowed sets from Dey \& Dey, 1984.

 The form of the potential is given below :
\begin{equation}
H_{i,j}~=~- \f{2\alpha_s\sigma^3}{3 m_i m_j
\pi^{1/2}}(\lambda_i.\lambda_j)(S_i.S_j)e^{-\sigma^2 r_{ij}^2}.
\label{diq}
\end{equation}
The factor $\sigma^3/\pi^{1/2}$ normalizes the potential. In this
equation $\alpha_s$ is the strong coupling constant, and the
subscripted $m, \lambda$ and $S$ are the masses, colour matrices
and spin matrices for the respective quarks.

   For u-d quarks Dey et al. (1984) found that this gives
$\sigma$ varying from 6 to 2.03 fm$^{-1}$ for a set of $\alpha_s$
0.5 to 1.12. The parameters we have used  are given in
Table~\ref{tabparam}.

It is found that diquark binding depends strongly on the strength
and range of spin-spin interaction which are interconnected via
hadron phenomenology. This is irrespective of whether it is
deduced from a the Fermi-Breit interquark force or an
instanton~-~like four fermion interaction as talked of, for
example, in Rajagopal et al.(2000).

\begin{table}
\caption{Parameters of the Gaussian Potential} \vskip 1cm
\begin{center}
\begin{tabular}{|c|c|c|}
\hline Sets & $\alpha_s$ & $\sigma (fm^{-1})$\\ \hline \hline 1 &
0.5 & 6.0 \\ 2 & 0.5 & 4.56\\ 3 & 0.87 & 6.0\\ 4 & 0.87 & 2.61\\ 5
& 1.12 & 6.0\\ 6 & 1.12 & 2.03\\ \hline
\end{tabular}
\end{center}
\label{tabparam}
\end{table}

\section{The effect of the potential on diquarks.}

Anti-symmetry  of   flavour  symmetric di-quark wave function
requires that while space part is symmetric, di-quark must be
either in spin singlet and colour symmetric ($6$) state, or in
spin triplet and colour anti-symmetric ($\bar 3$) state. In both
cases spin-spin force is repulsive \footnote{ Private
communication, R. K. Bhaduri.} and formation of pair is inhibited.

For flavour anti-symmetric  di-quarks, however,  the situation is
the opposite. Colour symmetric $6$ configuration is associated
with the spin triplet so that $(\lambda_i.\lambda_j)(S_i.S_j) =
1/3 $ and colour anti-symmetric state ($\bar 3$) goes with the
spin singlet which gives $(\lambda_i.\lambda_j)(S_i.S_j) = 2$.
With overall negative sign in the potential (\ref{diq}) these
channels produce attraction. Hence there is a probability for
example of u, d quarks to pair up predominantly in spin singlet
state. The effect of this can be found easily in our model since
we know the distribution of the u, d and s quarks in the momentum
space and their Fermi momenta are uniquely determined from precise
and lengthy calculations satisfying beta stability and charge
neutrality.

In addition to spin-colour contribution the potential
Eq.(\ref{diq}) is evaluated in the momentum space :
\begin{equation}
\f {1}{4\pi^3~ 3~ x \rho_0}\f{\alpha_s \sigma^2}{3m_im_j}\int
f(k) k_i^2 k_j^2 dk_i dk_j dcos(\theta)
\label{cor}
\end{equation}
where  $x \rho_0$ is the density at the star surface where the
energy per baryon is minimum (x = 4.586, 4.014 for eos1, eos3):
\begin{equation}
f(k)=\f{1 - {\rm exp} ( \f{-k^2}{\sigma^2})}{k^2}
\end{equation}
and
\begin{equation}
k^2 = \f{k_i^2 + k_j^2}{4}-\f{k_i k_j cos(\theta_{ij})}{2}
\label{relk}
\end{equation}

It is to be noted that Fermi momenta for u, d, s particles are
different. Thus the contribution of a specific  di-quark in the
energy can be simply the integral (\ref{cor}) and the colour spin
factor. Maximum contribution is around the Fermi surface though.
These are given in Table (\ref{corr}).

\begin{table}
\caption{Integrated values for the pairing energy Eq.(\ref{diq})
for different pairs for spin singlet (colour $\bar 3$) states in
MeV. For spin triplet (colour $6$) state the energies will be six
times less. } \vskip 1cm
\begin{center}
\begin{tabular}{|c|c|c|c|c|c|c|}
\hline EOS & Sets & $\alpha_s$ & $\sigma$&
\multicolumn{3}{c|}{diquark type}\\  \hline
 &&&$fm^{-1}$ & ud&ds&su\\
\hline eos1 & 1 & 0.5 & 6.0 & -3.84 & -1.45 & -1.23\\ & 2 & 0.5 &
4.56 &-3.79&-1.44&-1.22\\ & 3 & 0.87 & 6.0&-6.68&-2.53&-2.22\\ &4
& 0.87 &2.61&-6.22&-2.37&-2.02\\ &5 & 1.12 &
6.0&-8.59&-3.25&-2.76\\ &6 & 1.12 & 2.03&-7.59&-2.89&-2.48\\
\hline eos3 & 1 & 0.5 & 6.0 & -3.87 & -1.40 & -1.15\\ & 2 & 0.5 &
4.56 &-3.83&-1.39&-1.17\\ & 3 & 0.87 & 6.0&-6.74&-2.44&-2.06\\ &4
& 0.87 &2.61&-6.32&-2.29&-1.95\\ &5 & 1.12 &
6.0&-8.68&-3.14&-2.65\\ &6 & 1.12 & 2.03&-7.74&-2.82&-2.40\\
\hline
\end{tabular}
\end{center}
\label{corr}
\end{table}
Note that there is a difference between this energy and the
conventional pairing, where the effect of a long-range potential
is a shift which is found by solving the gap equation. This is
more like a correlation energy for some of the paired diquarks in
flavour anti-symmetric state.

The Table (\ref{corr}) shows that the variation of the correlation
energy is significant, when different sets for the smearing in
the spin-spin potential, are chosen. The variation with the
equation of state (EOS) eos1 and eos3 is comparatively unimportant
{\footnote {As stated before, these EOS differ only in one
parameter which controls the chiral symmetry restoration for the
quark masses at high density (D98)}}. We also see that the ud
pairing correlation energy is substantially larger than that of
the other pairs su and ds.

Let us recall that the energy per baryon is 888.8~ $MeV$ with eos1
and 844.6~ $MeV$ with eos3 as compared with 930.4 for $^{56}Fe$-
matter. We can see that even in the preferentially ordered spin
singlet state, one has only a few $MeV$ extra binding on the
average for every diquark, compared to a positive energy of
several hundred $MeV$.

However one should not forget that in a thermonuclear reaction
(TR) every fusion produces energy which is precisely of this
order. On the other hand TR is fast. To establish a stable high
density of about ~ 4.5 times $\rho_0$ and to get back the ordering
of the diquarks after a TR must take long time. If it is
established that the concerned stars are indeed strange stars and
the diquark pairing is the phenomenon responsible for long
lasting bursts, then one could claim a link between the smallest
quarks and the densest stars as has been pointed out before Ray
et al. (2000).

\begin{figure}
\psfig{file=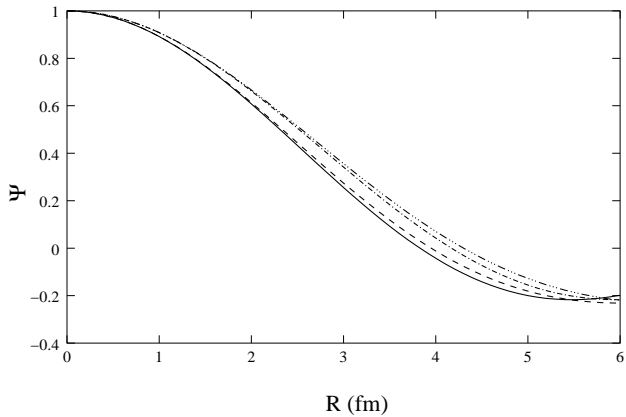,width=0.48\textwidth} \caption{The figure
shows the  similarity between the spherical Bessel function and
the appropriate oscillator wave functions.  The top pair of curves
correspond to $cos(\theta)~  = ~ 1$ in  Eq.(\ref{relk}) and the
bottom  curves to $cos(\theta) ~=~  - ~1$. The oscillator in both
cases is the  upper curve of the pair. The values  of relative $k$
are 0.75  and 0.82  $fm^{-1}$ respectively. At  such  a relatively
large momenta very little angular dependence is seen.} \label{osc}
\end{figure}

\section{Conclusions and summary.}

Our calculations teach us the following :

(1) There are antisymmetric diquark states for dissimilar quark
pairs in the spin parallel and antiparallel states with the
attraction six times stronger for the latter compared to the
former. But the magnitude of the attraction depends strongly on
the form of the interaction, even when the interaction is fitted
to observables like the standard isobar-nucleon mass difference.

(2) However, the six parameter sets that we have considered all
show an attraction of few $MeV$ so that it is comparable to other
strong interaction phenomenon like energy release per particle in
a thermonuclear burst. Since our model consists of realistic
strange stars with quarks at the surface and not in the interior
as in hybrid neutron stars, there is bound to be observable
surface phenomenon.

(3) The interaction producing a coloured diquark in spin zero
state, for example, is a strong one and its overall effect is
lowering of energy by $2$ to $7~MeV$. Once the pairs are
misaligned due high level accretion of some binary stars and
subsequent violent thermonuclear reactions (lasting typically for
$\sim 20~s$) their recombination may provide bursts over several
hours with energy release estimated to be large. The crucial fact
is that the recombination time scale is long, since the strong
interaction pairing process is supplemented by beta equilibrium
and charge neutralization which are slower weak and
electromagnetic processes. The number of pairs is shown to be
right to produce the estimated energy release for 4U~1820$-$30.

(4) The alternative to this calculation is to consider the full
16-component Dirac wave function for the diquark in a manner done
by Crater and van Alstine (1984) using the Dirac constraint method
for the two body Dirac equation. This is clearly beyond the scope
of the present paper which is concerned more with phenomenology.
In such a calculation the effect of the spin - spin force will be
manifest in the mean field level with more complicated spin wave
functions but we are not sure if such states can be used to
generate solutions of the TOV equations.

In summary we suggest that the superbursts (sometimes repeated),
lasting long hours, may be due to breaking of unlike quark pairing
in a specific coloured state in strange quark stars, following
conventional quick thermonuclear bursts and their subsequent
recombination. If strange stars are confirmed from
astro-phenomenology, such considerations may prove to be very
useful.

\section*{Acknowledgments}

The authors MS, SR and JD are grateful to IUCAA, Pune, India, for
a short stay. It is also our great pleasure to thank Prof. R. K.
Bhaduri and Mr. Ashik Iqubal for many discussions. We are
grateful to Dr. Siddhartha Bhowmick for careful perusal of the
manuscript.



\begin{thebibliography}{}
\bibitem[]{} Bhaduri R. K., L. E. Cohler \&  Nogami Y. 1980, Phys.
Rev. Lett. 44, 1369.

\bibitem[]{} Bhalerao R. S. \&   Bhaduri R. K. 2000,  `Droplet formation
in quark-gluon plasma at low temperatures and high densities',
hep-ph/0009333.

\bibitem[]{} Bildsten L., `Theory and Observations of Type I X-Ray
Bursts from Neutron Stars', 2000, astro-ph/0001135

\bibitem[]{}  Blinder S. M. 1980, J. Mol. Spec. 5, 17.

\bibitem[]{} Bombaci I. \& Datta B. 2000, ApJ 530, L69.

\bibitem[]{} Cornelisse R.,  Heise J.,  Kuulkers E.,  Verbunt F.
\&  in 't Zand J. J. M. 2000, Astron. \& Astrophys. 357, L21.

\bibitem[]{} Cornelisse R., Kuulkers E.,
in 't Zand \& Verbunt F. J. J. M.  and Heise J.,2002, Astron. \&
Astrophys. 382, 174.

\bibitem[]{} Crater H. W., \& van Alstine P. 1984, Phys. Rev. Lett.
53, 1527.

\bibitem[]{} Cummings A. and Bildsten L., 2001, Ap. J. 559, L 127.


\bibitem[]{} Dey J. \& Dey M. 1984, Phys. Lett.  B 138, 200.

\bibitem[]{} Dey M., Bombaci I.,  Dey J., Ray S. \& Samanta B.C.
1998, Phys.  Lett. B438, 123; Addendum 1999, B447, 352; Erratum
1999, B467, 303; 1999, Indian J. Phys. 73B, 377.

\bibitem[]{} Drake J. J. et al, ``Is RXJ~1856.5$-$3754 a Quark
Star", astro-ph/0204159 (2002).

\bibitem[]{} Heise J., in 't Zand  J. J. M.  \&  Kuulkers E. 2000,
AAS HEAD Meeting, 32, 28.03.



\bibitem[]{} Kapoor R. C.  \&  Shukre C. S. 2001, Astron. \& Astrophys.
375, 405 ( astro-ph/0011386).

\bibitem[]{} Kapusta K. 1981, Phys. Rev. D23,  2444.

\bibitem[]{} Kuulkers E., Homan J., van der Klis, M., Lewin W. H.
G. \& M$\acute{e}$ndez M. 2002, Astron. \& Atrophys. 382, 947.
(astro-ph/0105386).

\bibitem[]{} Kuulkers E., in 't Zand J. J. M., van Kerkwijk M. H., Cornelisse
R., Smith D. A., Heise J., Bazzano A., Cocchi M., Natalucci L. \&
Ubertini P. 2002b, Astron. \& Astrophys. 382, 503.
(astro-ph/0111261)

\bibitem[]{} Li X., Bombaci I., Dey M., Dey J. \& van den Heuvel
E. P. J. 1999a, Phys. Rev. Lett. 83, 3776.

\bibitem[]{} Li X., Ray S., Dey J., Dey M. \& Bombaci I. 1999b,
ApJ 527, L51.

\bibitem[]{} Rajagopal K. \&  F. Wilczek 2000,  `The Condensed
Matter Physics of QCD', hep-ph/0011333, to be published in
Festschrift in honour of B. L. Ioffe.

\bibitem[]{} Ray S., Dey J. \& Dey M. 2000, Mod. Phys. Lett. A15,
1301.

\bibitem[]{} Strohmayer T. E.  \&  Brown E. F. 2001, `A remarkable three
hour thermonuclear burst from 4U~1820$-$30', astro-ph/0108420.


\bibitem[]{} Wijnands R. 2001, Ap. J. 554, L59.

\bibitem[]{} Wijnands R. et al.  2001, Ap. J. 560, L159.


\bibitem[]{} Wijnands R. et al., The erratic luminosity behaviour
of  SAX~J1808.8$-$3658 during its 2000 outburst,  2001b,
astro-ph/0105446, Ap. J. (in press).


\end{thebibliography}
\end{document}